# Shear relaxation behind the shock front in ⟨110⟩ Molybdenum - From the Atomic Scale to Continuous Dislocation Fields

Roman Kositski[1], Dominik Steinberger[2], Stefan Sandfeld[2], Dan Mordehai[1]

[1] Department of Mechanical Engineering, Technion – Israel Institute of Technology, 32000, Haifa, Israel
[2] Chair of Micromechanical Materials Modelling (MiMM), Institute of Mechanics and Fluid Dynamics, TU Bergakademie Freiberg, 09599 Freiberg, Germany.

Corresponding author: Roman Kostiski. Email: kositski@gmail.com.

## Abstract

In this work we study shock-induced plasticity in Mo single crystals, impacted along the <110> crystal orientation. In particular, the shear relaxation behind the shock front is quantitatively inspected. Molecular dynamics (MD) simulations are employed to simulate the deformation during shock, followed by post-processing to identify and quantify the dislocation lines nucleated behind the shock front. The information on the dislocation lines is ensemble averaged inside slabs of the simulation box and over different realizations of the MD simulations, from which continuous dislocation fields are extracted using the Discrete-to-Continuous method. The continuous dislocation fields are correlated with the stress and strain fields obtained from the MD simulations. Based on this analysis, we show that the elastic precursor overshoots the shear stress, after which dislocations on a specific group of slip planes are nucleated, slightly lagging behind the elastic front. Consequently, the resolved shear stress is relaxed, but the principal lateral stress increases. The latter leads to an increase in the resolved shear stress on a plane parallel to the shock wave, resulting in an additional retarded front of dislocation nucleation on planes parallel to the shock front. Finally, the two-stage process of plasticity results in an isotropic stress state in the plane parallel to the shock wave. The MD simulation results are employed to calculate the dislocation densities on specific slip planes and the plastic deformation behind the shock, bridging the gap between the information on the atomic scale and the continuum level.

## 1. Introduction

Shock deformation of metals was studied extensively in the past century as it is fundamental to the understanding of how metals behave at high strain rate. For instance, when a high-velocity projectile impacts a metallic target, pressures of the order of 10 GPa and strain rates of $10^4$ $s^{-1}$ can evolve at the impacted site [1]. A micro-meteorite can impact a satellite or a space station at velocities exceeding 10 km $s^{-1}$ resulting in pressures of a few tens of GPa and strain rates of $10^5$–$10^6$ $s^{-1}$ [2]. In explosion welding (or bonding) pressures of 20–30 GPa can be generated with high strain rates and significant plastic deformation at the joined interface [3]. Even in plane and car crashes, high pressures and high strain rates with large plastic deformation can occur [4]. Despite the considerable amount of research, only recently experiments [5–8] and numerical simulations [9–11] were able to shed some light on the microstructural origins of plastic deformation in metals during the initial stage of shock compression.

The most common technique to experimentally study shock-induced plasticity is by generating a planar shock on a surface of the specimen. As the planar shock wave propagates from the surface into the bulk, the material is compressed uniaxially. As a result, the state of strain right behind the shock is uniaxial with a three-dimensional stress state (stress components on the plane non-parallel to the shock plane are not zero). This stress state may result in shear stresses that are sufficiently high to cause plastic flow. During plastic deformation, the shear stresses are reduced and the stress state approaches an isotropic compression. This process of plastic strain evolution and concomitant reduction of shear stresses is known as plastic relaxation or shear relaxation [12].

Several works attempted in recent years to study the underlying microstructural mechanisms during plastic relaxation. Wehrenberg et al. [8] performed laser-driven shock experiments on single crystal Ta samples to pressures of up to 100 GPa. Using Laue X-ray diffraction they measured the strain anisotropy right behind the shock wave and could measure the relaxation time. They found that for pressures of about 50 GPa the relaxation time is of the order of 1 ns. For a shock pressure of 100 GPa, they could not observe the relaxation and concluded that it must occur in less than 0.3 ns, which is the temporal resolution of the experiment. Milathianaki et al. [5] performed similar experiments on single crystal Cu at pressures of about 73 GPa and measured the strain evolution *in-situ*. From their measurements, they concluded that the shear relaxation process takes about 60 ps with plastic strain evolving to about 6.2 %.

The experiments suggest that there is a substantial difference in relaxation times between lower and higher shock stresses. In order to rationalize this observation, a number of models for shear stress relaxation have been developed. Rudd et al. [13] proposed a model for calculating the stress at a given strain, based on the evolution of dislocation density and dislocation glide. With this model, they calculated the relaxation time in shock compressed Ta and found that for shock pressures of 10 GPa to 50 GPa the relaxation time is of the order of a few ns. In this pressure range, the dislocation density first needs to sufficiently increase so that the glide of dislocations can reduce the shear stresses. Motivated by the fact that their model does not account for homogeneous dislocation nucleation, they additionally performed molecular dynamics (MD) simulations. From these simulations, they found that for shock pressures above 66 GPa, dislocations homogeneously nucleate and the relaxation time drops by three orders of magnitude to about 1 ps. This significant difference in relaxation times suggests that the above mentioned experimentally relaxation times result from different dislocation mechanisms.

While the model by Rudd et al. captures the relaxation time to some extent, they employ a continuum model with phenomenological rules for the evolution of dislocation. In addition, homogenous nucleation is not taken into account. To study the plastic relaxation with details of the dislocation structure, and also to include homogenous nucleation, Shehadeh and Zbib [14] used a Discrete Dislocation Dynamics (DDD) framework to perform simulations of shock loaded single crystal Cu. Performing simulations above the nucleation threshold with shock compressive stresses ranging from 20 GPa to 100 GPa they found that the plastic relaxation time is decreasing from about 500 ps to about 150 ps as the shock intensity (and subsequently the strain rate) increases. Based on their DDD simulations they were able to correlate the evolution of dislocation densities to strains and stresses. However, despite the level of details in DDD simulations, their analysis relied on effective quantities, which are insufficient to understand the micromechanical mechanisms behind plastic relaxation in single crystals. In addition, reliable rules for homogeneous nucleation in DDD simulations are still lacking, and the role of homogenous nucleation in the stress relaxation is not clear.

MD simulations are more commonly used than DDD simulation and have been employed to study the plastic deformation behind the shock front and in particular the shear relaxation process. MD simulations naturally incorporate dislocation nucleation and the interaction and evolution of dislocations. In recent years, a large amount of work has been carried out on Face-Centered Cubic (FCC) metals [6,9,15,16] while simulations of Body-Centered Cubic (BCC) metals are less common. Most of the work on BCC metals was on Fe due to its relevance to technological applications. These works were focused on the solid-solid phase transformation occurring in Fe at about 13 GPa and thus less focused on dislocation–mediated plasticity [11,17–19]. For instance, in MD simulations by Cuesta-Lopez and Perlado [20] of Ta, W and Fe, solid-solid phase transformations were observed during shock, which prohibited the nucleation of dislocations, although only Fe was expected to deform via phase transformation. Liu et al. [21] performed MD simulations of shock loading of a single crystalline W specimen at different orientations. They obtained detailed stress profiles along the propagation of the shock front and observed the relaxation of the shear stresses due to plasticity behind the plastic shock front. While focusing on the evolution of stresses, they did not study the underlying micromechanical mechanisms causing the shear relaxation, nor performed a quantitative discussion on it. More recently Ravelo et al. [22] developed a new interatomic potential for Ta. They performed shock propagation MD simulations in single crystal Ta and found that at a certain pressure (or particle velocity) threshold, twins nucleate behind the plastic wave front reducing the shear stresses to practically zero. Tramontina et al. [23] used the same potential but added a nanovoid inside the lattice to act as a nucleation site for dislocations. They found that around the nanovoid dislocations nucleate to a dislocation density of about $10^{13}$ $cm^{-2}$ at shock pressures of about 50 GPa.

These examples also demonstrate the shortcomings of MD simulations; they hold temporal information on all atoms in the system and if one wants to relate the discreet nature at the atomic level to the continuum level, the information from all atoms should be related quantitatively to continuous fields [24]. For instance, while dislocations are discrete defects in the crystal lattice, their collective motion results in a continuous plastic deformation. On the other hand, dislocation-based continuum models of plasticity are not restricted to the particular details of each dislocation, as they describe dislocations as continuous fields on different slip planes with some physical rules from length scales below the averaging volume size. In most cases, the way the continuum is related to the atomic scale is by transforming individual dislocations from the MD simulations into a total dislocation density field [9,25]. However, such a field variable is insufficient for computing the spatially heterogeneous plastic deformation and the evolution of shear relaxation since different slip systems contribute differently to plastic deformation [26]. Thus, we would like to benefit from the advantages of MD simulations, in naturally describing dislocation mechanisms, and quantitatively relate them to continuous parameters of the plastic deformation.

One strategy would be to employ MD simulations to study shear relaxation behind the shock front and to develop computational frameworks for data mining that could help to bridge between the atomic and continuum scales. For instance, ensemble averaging and statistical data analysis of information on individual dislocation can be related in a sequential multiscale approach to continuous dislocation fields, in order to develop constitutive rules with quantitative information on the dislocation microstructural evolution and its interplay with the loading conditions. The challenges to fulfill this strategy is two-fold: first, it is necessary to extract quantitative information of dislocations from MD simulations. Recently developed tools, such as the Dislocation Extraction Algorithm (DXA) [27]) made it possible to obtain dislocations as geometrical, line-like objects from the atomic scale. Second, a well-defined strategy to obtain continuous dislocation fields from discrete dislocation lines is required for a suitable data analysis. This Discrete To Continuum (D2C) strategy was recently developed and has been applied to and validated by Discrete Dislocation Dynamics (DDD) simulations [28,29] and has been used for studying nano-scratching with MD simulations [30].

In this work, we employ the combined DXA-D2C strategy to study the evolution of continuous dislocation fields during plastic relaxation in shocked single crystal [110] Mo from MD simulations. In Sec. 2 we detail the simulation methods and, in particular, the method to transform discrete dislocation lines to continuous fields. In Sec. 3 we describe the MD simulation results and the outcome of extracting continuous dislocation. Based on these results, the correlation between continuous dislocation fields and stress that develops in the plastic zone behind the shock is discussed in Sec. 4, and the highlights of this paper are summarized in Sec. 5.

## 2. Methods

### 2.1. Molecular dynamics

We performed nonequilibrium MD simulations using the open-source code LAMMPS [31]. The interactions between the Mo atoms were described using the embedded-atom method (EAM) interatomic potential, with a parameterization proposed by Ackland and Thetford [32]. This potential was previously used in studying dislocation mobility in Mo [33] and dislocation nucleation and multiplication during plastic deformation of nanopillars [34]. While these works support the validity of this potential to describe plastic deformation, we are not aware of any shock simulation performed with the potential. To further validate the potential we first performed MD simulations using a Hugoniostat integration [35]. Using this method, we generated a simulation-based Hugoniot curve satisfying the Rankine-Hugoniot jump conditions for steady shocks. A single crystalline perfect lattice was compressed either by hydrostatic pressure conditions or by uniaxial deformation in one of the three low index directions: [100], [110] and [111]. In Figure 1 the calculated pressure vs. density curve is compared with the Hugoniot obtained from gun and laser experiments ([36] and reference therein). It can be seen that for pressures above 200 GPa the simulations start to noticeably deviate from the experimental curve. Thus, in the following sections we will limit our simulations and discussion to the regime below 200 GPa.

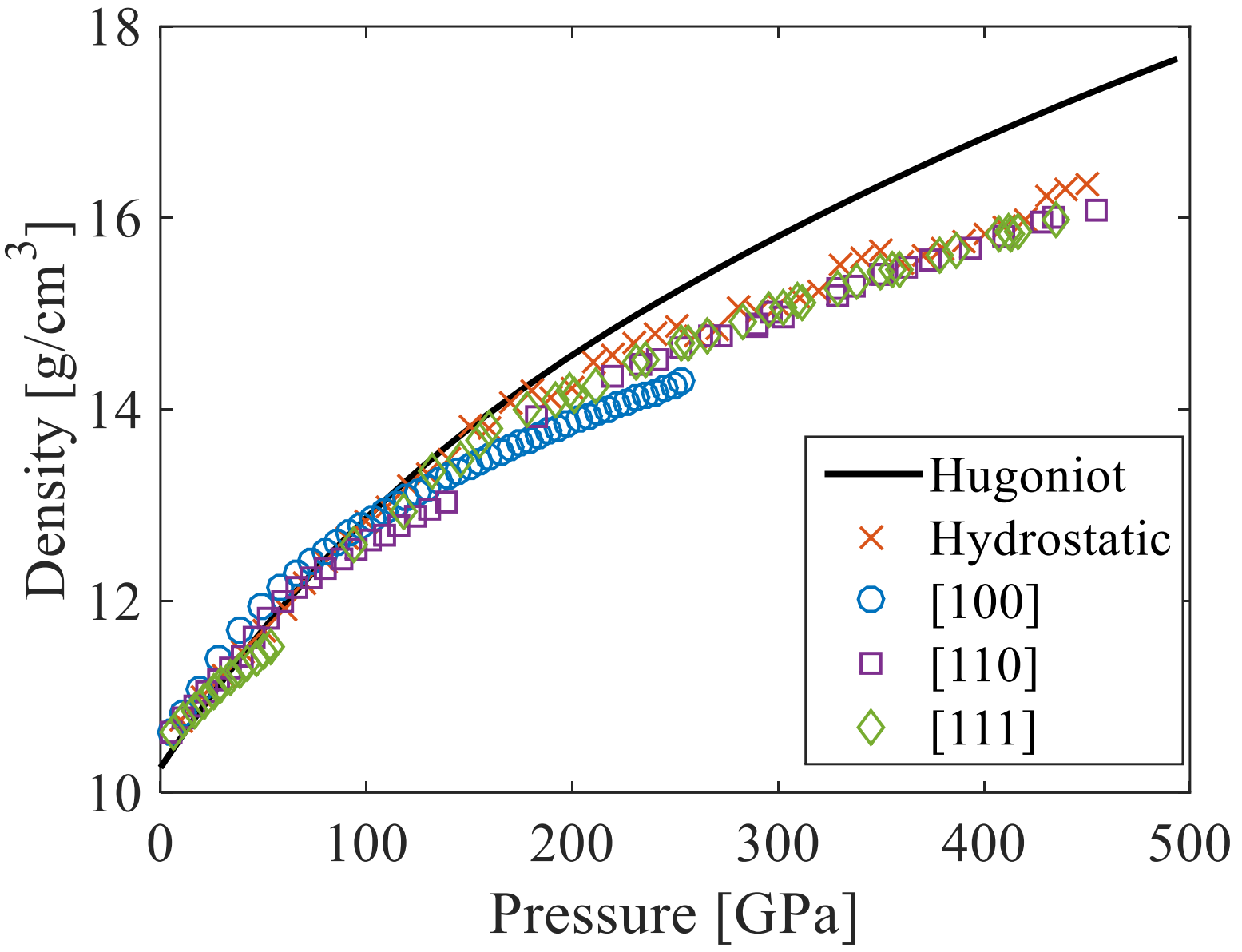

**Figure 1 - Hugonit curves for Mo obtained from simulations compared to experimental results** [36]**.**

In the rest of the paper we discuss MD simulations of shock propagation in Mo single crystalline thin-film that were impacted in the [110] direction (z-axis). Periodic boundary conditions along the x ([00$\bar{1}$] direction) and y ([$\bar{1}$10] direction) axes were imposed. Periodicity of 50 unit cells in the x and y directions (15 nm and 22 nm in the x- and y-axis, respectively), and a thickness of 300 unit cells (135 nm) along the z-axis were chosen, resulting in a total of about 3M atoms. In a recent study, it was found that these lateral dimensions are enough to obtain smooth shock profiles and reduce the effect of periodicity [22]. The simulation setup is shown in Figure 2. The sample was initially set to a temperature of 300 K and given sufficient time to relax to a stress-free state using a Nosé-Hoover thermostat and the Parrinello-Rahman barostat to approximate a NPT ensemble. Following the relaxation, the cell dimensions were fixed in the lateral (x-y) directions. These boundary conditions effectively model the uniaxial strain conditions occurring at the center of a specimen in plate impact experiments [37,38]. Shock loading of the sample was achieved by using a moving virtual piston. The piston's velocity was ramped from zero to the final velocity in a time range of 1-2 ps to reduce artificial heating of the interface [22].

During the MD simulation, the atom positions are stored and virial stresses for individual atoms, as well as mean values for slabs of atoms along the “z” direction, are computed at constant time steps. We note that in the virial stress calculation of a group of atoms, the compressed volume is considered, based on Voronoi tessellation analysis of the volume around each atom. We use the OVITO code [39] with its built-in DXA [27] to identify dislocation lines and their Burger vectors. The individual dislocation lines are transformed into continuous dislocation fields using the D2C analysis described in the following.

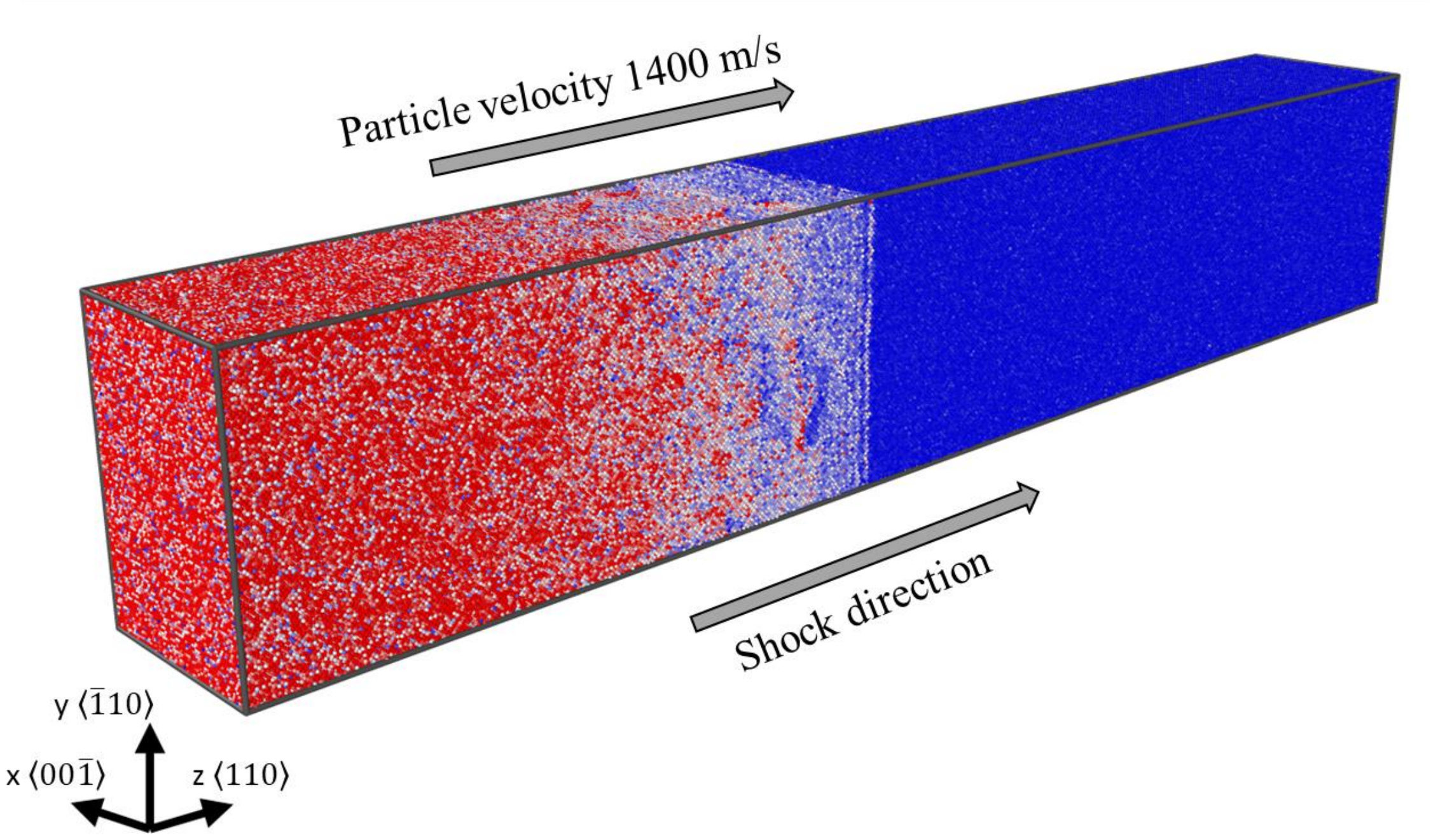

**Figure 2 - The atomic configuration in the MD simulation. The shock direction is along the <110> direction (z-axis). Atoms are colored according to their velocity, with red being 1400 m s$^{-1}$ and blue is zero velocity.**

## 2.2. Discrete to Continuum (D2C)

Dislocations form highly complex microstructures behind the shock front. Analyzing the dislocation configuration and relating it to the mechanical response in its discrete form proves challenging because of the large number of dislocations involved. Also, due to the size limitation of MD simulations, the microstructure of one simulation might not capture the whole set of features possible under these conditions, and it may require several simulations of the same ensemble to reveal the average microstructure response of the system.

A remedy is a description of the dislocation microstructure as continuous fields, i.e., dislocation densities. The most naive approach to calculating the dislocation density would be to compute the total length of dislocation lines in a simulation box and divide by the box volume. This method was previously adapted in MD simulations [25] but a significant drawback of this approach is that one can only obtain a single value evolving with time, assuming the dislocation structure is homogeneous throughout the specimens. All local features are hence lost in the conversion process.

Within the D2C method introduced in [27–29] dislocation lines are converted into continuous fields while taking their local features into account, both by spatial discretization, and employing a special set of fields. In the following, the principles behind this method are outlined.

A dislocation may be fully described as a parameterized, oriented curve $\boldsymbol{c}(u)$ with $u \in [0, 1]$ with Burgers vector $\boldsymbol{b}$. The local character of the dislocation, i.e., whether it is of screw, edge, or mixed type, depends on the angle $\varphi$ between $\boldsymbol{b}$ and the line tangent $\boldsymbol{l}(u)$. This dependence of the dislocation character on the local orientation of the dislocation line with respect to its Burgers vector needs to be reflected in the continuous description. The field variables of the Continuum Dislocation Dynamics (CDD) theory [39-41] are well suited as such descriptor. Instead of attempting to capture the orientation dependence of the dislocations in several dislocation density types, Hochrainer introduced the orientation as an additional dimension in CDD and is thus able to capture all characteristics of a dislocation in a higher-dimensional dislocation density $\rho(\boldsymbol{r}, \varphi)$, with $\boldsymbol{r}$ being the spatial coordinate. By expanding $\rho(\boldsymbol{r}, \varphi)$, an infinite hierarchy of dislocation densities can be introduced [42]:

$$\boldsymbol{\rho}^{(n)}(\boldsymbol{r}) = \int_0^{2\pi} \rho(\boldsymbol{r}, \varphi)\boldsymbol{l}(\varphi)^{\otimes n} \, \mathrm{d}\varphi. \tag{1}$$

Here, $\boldsymbol{l}(\varphi)^{\otimes n}$ denotes the $n$-times outer product of the line vector $\boldsymbol{l}$ associated with the aforementioned angle $\varphi$. The amount of information contained within the continuous description can thus be adjusted by truncation of the series. For example, the zeroth order density $\rho^{(0)}(\boldsymbol{r})$ obtained in this manner represents the total dislocation density. The first order density $\boldsymbol{\rho}^{(1)}(\boldsymbol{r})$ is a first-order tensor and denotes the excess dislocation density, i.e., the density of so-called geometrically necessary dislocations. Information about the character-dependent density, regardless the sign of the line orientation, is recovered when taking the second order density tensor $\boldsymbol{\rho}^{(2)}(\boldsymbol{r})$ into account. Thus, in a robust manner, more and more information about the system can be taken into account, if necessary.

For the numerical computation of the fields, space is discretized into sub-volumes $\Omega_i$. The different densities are then computed in each sub-volume via

$$\boldsymbol{\rho}_{\Omega_i}^{(n)} = \frac{1}{V_{\Omega_i}} \sum_{c \in \Omega_i} \int_{\mathcal{L}_{\Omega_i}^c} \boldsymbol{l}\big(\varphi(u)\big)^{\otimes n} \, \mathrm{d}u, \tag{2}$$

with $V_{\Omega_i}$ denoting the volume of the respective sub volume $\Omega_i$, and $\mathcal{L}_{\Omega_i}^c$ denoting the part of curve $c$ contained within the sub volume $\Omega_i$.

Another field variable providing information about the dislocation configuration in an averaging volume is the so-called Kröner–Nye tensor [40,41]

$$\boldsymbol{\alpha} = \frac{1}{V_{\Omega_i}} \sum_{c \in \Omega_i} \int_{\mathcal{L}_{\Omega_i}^c} \boldsymbol{b}_c \otimes \boldsymbol{l}\big(\varphi(u)\big) \, \mathrm{d}u, \tag{3}$$

which is a measure for the plastic distortion due to the presence of dislocations. If the coordinate system in which the tensor components are computed is aligned with the Burgers vector and the direction of edge dislocations, the 11 and 12 tensor components are related to the geometrically necessary densities of edge and screw components.

The resulting fields can then be used, e.g., to compare different realizations of the same system with different initial velocities in the case of molecular dynamics, via subtraction of the fields. Another benefit is the possibility of computing “ensemble averages,” i.e., the average dislocation microstructure of different realizations of the same ensemble, by simple mathematical averaging of the fields. This enables the interpretation of the average response of a system instead of dealing with the peculiarities some systems may exhibit due to “special” initial and/or boundary conditions.

## 3. Shock-Induced Plasticity in ⟨110⟩ Mo

### 3.1. Molecular dynamics simulation results

Mo thin-films were impacted at a velocity of 1400 m $s^{-1}$. This velocity, as well as all impact velocities considered in this work, is above the Hugoniot Elastic Limit (HEL) and the maximum pressures are below 200 GPa. Moreover, very limited, if any, phase transformation is induced far from the shock front, i.e., plasticity is almost solely dislocation-mediated. In Figure 3a we show compressive stress profiles along the shock direction ($\sigma_z$) at t=4, 9, and 14 ps, after initial contact of the piston with the sample. The average particle velocities and virial stress tensor are calculated for slabs of atoms of thickness of 0.7 nm in the z direction. We define stresses to be positive if they are compressive stresses. As the piston hits the sample, there is an abrupt rise compressive stress/particle velocity followed by a gradual increase up to steady state values. Generally speaking, the abrupt stress jump can be ascribed to the elastic precursor, followed by a plastic shock front. However, in order to better understand the evolution of the stress profile behind elastic precursor, the evolution of the dislocation structure is analyzed.

The analysis of the evolution of the dislocation structure is demonstrated in Figure 4 for a specific atomic configuration. Using common neighbor analyses (CNA) and the DXA, a complex dislocation structure is identified within the plastically deformed region. We do not show here the nucleation process, but we rely on the previous MD simulations that showed that loops are nucleated behind the shock front [42,43]. The nucleated dislocations quickly expand and interact, transforming into a 3-dimensional dislocation structure. Qualitatively, two regions of dislocation structures can be identified which have two different densities, with the denser region appearing further away from the shock front. Also, a few small cells of Hexagonal Closed-Pack (HCP) phase can be seen. Since Mo is expected to retain its BCC structure up to pressures of 380 GPa, we believe the small HCP cells are an outcome of the interatomic potential [36]. Nonetheless, since those cells are small and are significantly behind the wave front and in the region where the stress gradients have diminished, their contribution to the plastic response right behind the shock front is considered here as negligible.

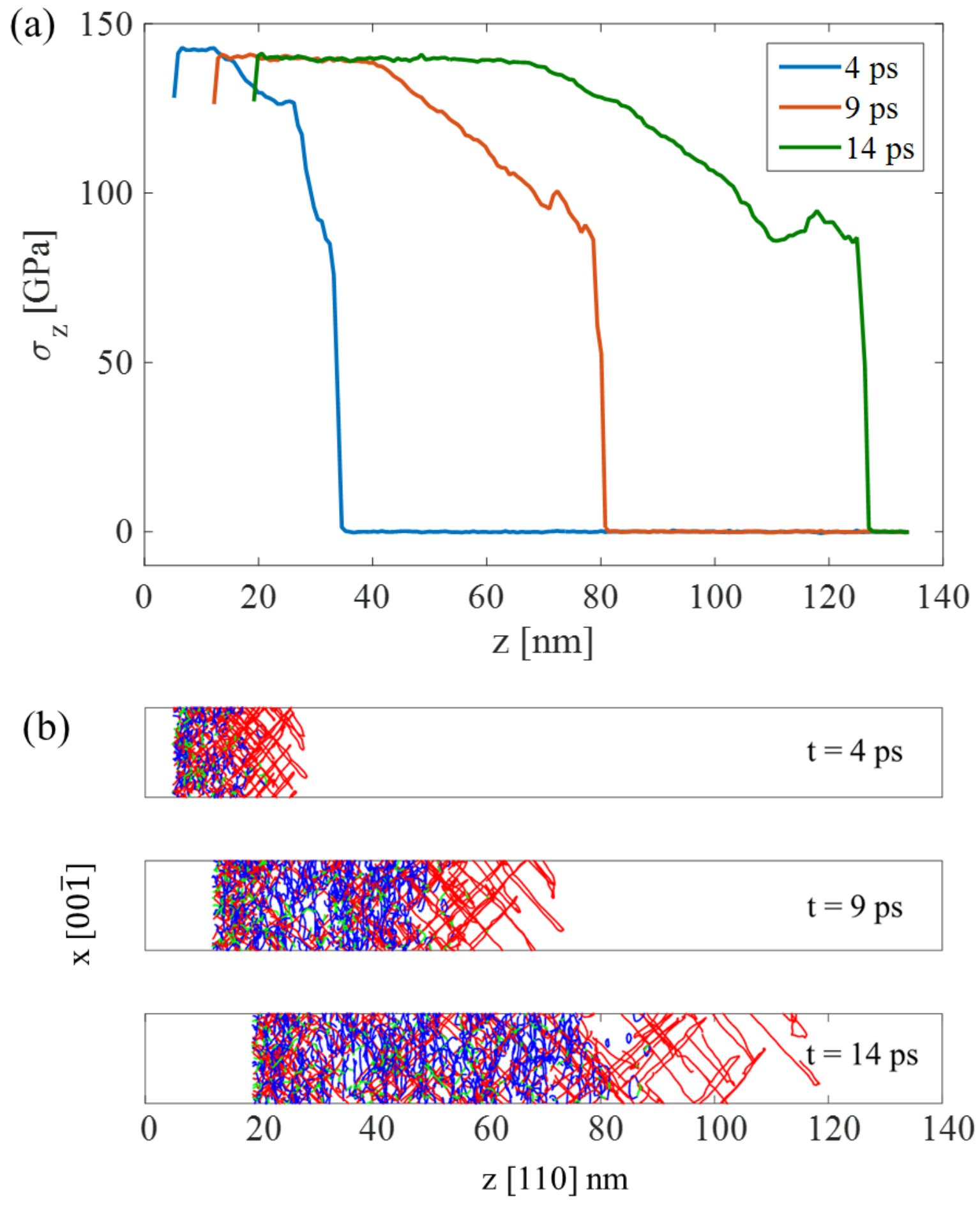

**Figure 3 - (a) Average compressive stress profiles along the sample at 4, 9 and 14 ps. (b) Dislocations in the sample viewed on the x-z plane. Out-of-plane and in-plane dislocations are colored in red and blue, respectively. The remaining dislocations are colored in green.**

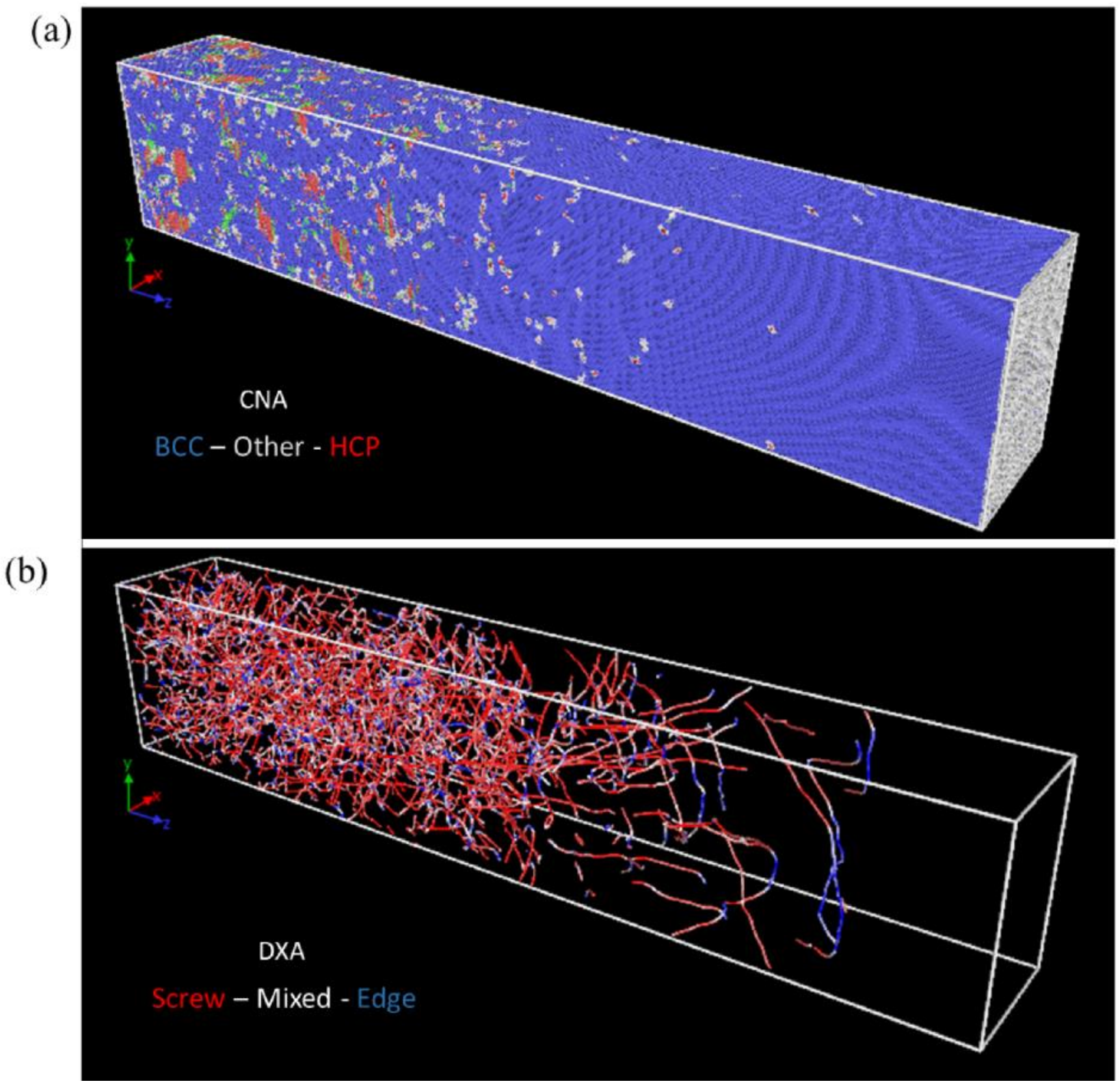

**Figure 4 – The dislocation structure formed during shock. (a) CNA applied to the atomic structure at t=14 ps. Blue shaded atoms are BCC structure, red are HCP phase and gray are dislocations or other lattice defects. (b) Dislocation lines calculated with the DXA algorithm and colored based on their character; red for screw and blue for edge segments.**

Let us first focus our discussion on the less dense dislocation region behind the shock front. Using the DXA algorithm we can categorize dislocations by their slip systems. The nucleated dislocations in the less dense region are residing on two possible slip systems: ½$[111](11\bar{2})$ and ½$[11\bar{1}](112)$. The Burgers vectors of these dislocations are in the x-z plane at an angle of 35° to the shock front. Being on a plane at an angle to the shock plane, we name them as “out-of-plane” dislocations. In Figure 3b we plot 2D projections of the extracted dislocation lines into the x-z plane. The dislocations in the out-of-plane slip systems are colored in red in Figure 3b. It can be seen from this figure that dislocations were nucleated in the out-of-plane direction and created a "slab" of a dislocation network with segments in the out-of-plane directions. The dislocations within this region have a significant screw component. One reason is that in this configuration Peach-Köhler forces acting on the edge segments to expand the loop are higher. Additionally, edge segments have a higher mobility [44], which makes the edge component glide more rapidly leaving behind “stems” of long screw segments. The width of the slab of out-of-plane dislocations expends as the shock wave propagates owing to the rapid glide of the leading edge components and the nucleation of new dislocation loops. Following the out-of-plane dislocations region, dislocations with two other Burgers vectors, ½$[1\bar{1}1]\{110\}$ and ½$[1\bar{1}\bar{1}]\{110\}$, are nucleated at the back of the slab. These slip systems lay on the $x - y$ plane parallel to the shock front, thus named “in-plane” dislocations. These dislocation lines, colored in blue in Figure 3b, are attributed to the denser region of the dislocations behind the shock front. The late appearance of the in-plane dislocations leads us to believe that their nucleation is stimulated by the existence of the out-of-plane dislocations, in addition to the compressive external stress induced by the impact. We propose that the conditions to nucleate the in-plane dislocations determine the thickness of the out-of-plane dislocations slab. While we shall further discuss these conditions in Sec. 4.1, we provide here a qualitative rationalization for the delayed nucleation of the in-plane dislocation, using the stress profiles.

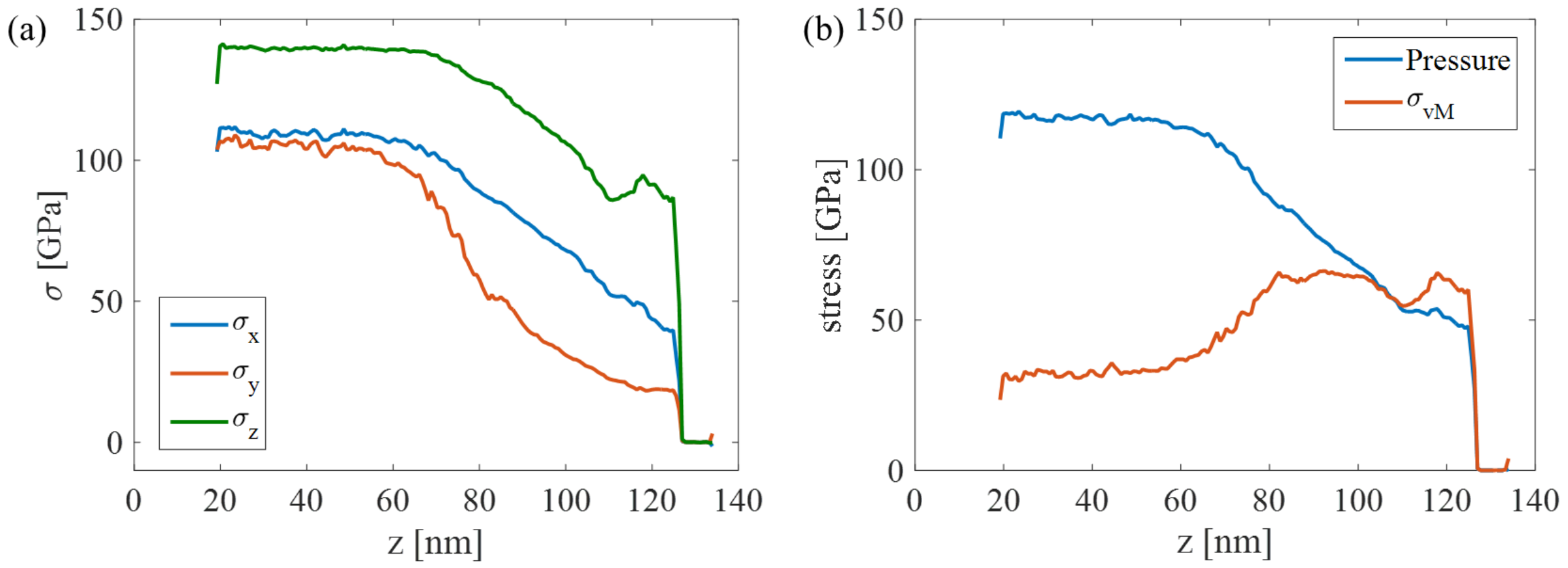

**Figure 5 – Stress profiles along the sample at 14 ps after initial loading. (a) The principal stresses. (b) The pressure and von-Mises stress**

In order to explain why the in-plane dislocations appearance is retarded behind the out-of-plane dislocations we recall that the plasticity behind the shock front is related to the stress profiles, which in the experimental literature is either based on the uniaxial stress or the pressure. However, the difference between them is actually very important: in a purely hydrostatic stress state there is only pressure, i.e. all principal stresses are the same, and there is no shear stress. But understanding shock-induced plastic deformation is a different situation since during shock loading the principal stresses are generally not equal. More particularly, the deformation behind the shock front rather corresponds to uniaxial strain. Only if the sample can sustain low shear stresses, compared to the compressive stresses, then the values of uniaxial stress and pressure are very close and one needs not to differentiate between them. But if the sample’s strength is relatively high, as in the case of perfect single crystals, one should consider the full stress tensor, or equivalently the three separate principle stresses, instead of just the pressure. For this reason, and in order to understand the nucleation of dislocation in the out-of-plane and in-plane directions, we examine the principal stress profiles during compression.

Figure 5 shows the principal stresses $\sigma_x$, $\sigma_y$ and $\sigma_z$, the pressure $P$ and the von-Mises stress $\sigma_{vM}$ profiles at a time of 14 ps. In both the elastic and plastic wave fronts, stress anisotropy can be observed; $\sigma_z$ is the highest stress, as expected from the stress in the loading direction, followed by $\sigma_x$ and the lowest principal stress $\sigma_y$. $\sigma_z$ rises abruptly at the elastic shock front to a value of about 80 GPa. Owing to the uniaxial-strain conditions imposed in the simulation, $\sigma_x$ and $\sigma_y$ also rise to values of 40 GPa and 18 GPa behind the front, respectively. The different values in the lateral $x$ and $y$ directions are attributed to the anisotropic elastic response of the crystal. Some minor variation in the stress components in the $x$ and $z$ directions are found within the elastic region, which will be neglected in the following discussion. Once the out-of-plane dislocations are nucleated, all principle stresses are found to deviate, with a generally increasing trend. When dislocations in the in-plane directions are nucleated (see the left part of the curves), these shear stresses are relived, which results in both principal stresses, $\sigma_x$ and $\sigma_y$, having very close values. Indeed, far from the shock, the lateral stress components reach plateaus of approximately 104 GPa and 108 GPa for $\sigma_x$ and $\sigma_y$, respectively. Thus, the nucleation of the dislocation structure behind the shock front on different slip systems contributes to the isotropic stress state in the shock plane (x-y plane). As a result, the von-Mises stress in Figure 5b, calculated from the three principal stresses, is reduced in what is known as “shear relaxation” [13,14]. While the shear stresses are reduced the pressure increases until reaching a steady state of 117 GPa for this case. In order to better relate the stress profiles and the dislocation structures formed behind the shock front, we employ the D2C analysis method on the MD simulation results, to transform the MD-obtained individual dislocation lines into continuous dislocation fields.

### 3.2. Continuous dislocation fields evolution

To better understand the evolution of the 3D stress state behind the shock front, we transform the dislocation lines into continuous dislocation fields as a function of position and time, maintaining information for each slip plane. Using the dislocation lines obtained from the DXA analysis and processing the data with the D2C method, we calculated the evolution of dislocation density fields on individual slip systems. The conversion of discrete lines to these fields requires the orientation $\varphi$ of the dislocation line in a fine manner. We therefore fit splines to the nodal description returned from the DXA algorithm. Averaging over the values in voxels that share the same $z$ coordinate, we obtain 1D plots of the dislocation density.

In Figure 6 we show the evolution of dislocation density profiles in the out-of-plane and in-plane directions. It can be seen that the dislocation density increases from zero to about $2.5\times10^{13}$ cm$^{-2}$ far from the shock, where the evolution of dislocations densities reaches a steady state value. Due to the discrete nature of dislocation lines and the limited size of our simulation cell, the density fields in Figure 6 are not smooth. In order to reduce the noise in the curves and to obtain smoother profiles, which we would have expected to get if the later size was substantially larger, we have repeated the simulation five times, with different initial atomic velocities (but for the same initial temperature). Ensemble averaging the dislocation density fields over the five realizations resulted in smoother curves, which are less affected by the limitation of our simulation box size (we defer further discussion on this to Sec. 4.1) .

The dislocation density contributions of lines along the real space coordinates, as well as the Kröner–Nye tensor components for an average of five realizations at a time of 14 ps after the impact, are shown in Figure 7. The components of the Kröner-Nye tensor are shown with respect to the $x$-$y$-$z$ sample coordinate system. There, e.g., $\alpha_{xy}$ corresponds to a plastic distortion in $x$-direction caused by line segments with net line orientation parallel to the $y$-direction.

Taking a look at the total density components in all three spatial directions (Figure 7 top) we observe that directly at the shock front, the main density contribution stems from dislocation lines parallel to the $y$-direction ($\rho_{yy}^{(2)}$, orange line), followed by an increase of contributions of the $x$- and $z$-line-directions. Once the density reaches a plateau ($z < 75$ nm) all contributions are roughly equal except for a slightly increased contributions from the $y$-line-direction. Since the plots are the result of averaging over a number of realizations we conclude that this is generic behavior, which is to be expected to be found in every sample. The reason for the asymmetric behavior in $x$- and in $y$-directions is the asymmetry in the slip planes in the out-of-plane directions (there are only two planes, inclined in one of the directions).

The components of the Kröner–Nye tensor corresponding to line segments parallel to $x$ directions (Figure 7, second from top) and parallel to the $z$ directions (Figure 7, bottom) fluctuate slightly around zero at the dislocation front at the right. While the components in both sub-figures also tend to fluctuate around zero in the region of the density plateau, the fluctuations of the $x$-line-direction components are more pronounced (Figure 7, second from top, $20\text{ nm} \leq z \leq 90\text{ nm}$). Although the components for the line directions parallel to $y$ (Figure 7, second from bottom) also fluctuate around zero in the density plateau, the shock front exhibits distinct features: at the front (around $z = 120$ nm), a concentrated dislocation structure in positive $z$-direction (green line) are accompanied by those in negative $x$-direction (blue line). The latter contribution stays constant until the fluctuations approaching the density plateau begin. The former, however, drops to zero and spikes again just before the fluctuations. Components corresponding to the contribution to a distortion in $y$-direction remain zero in this regime.

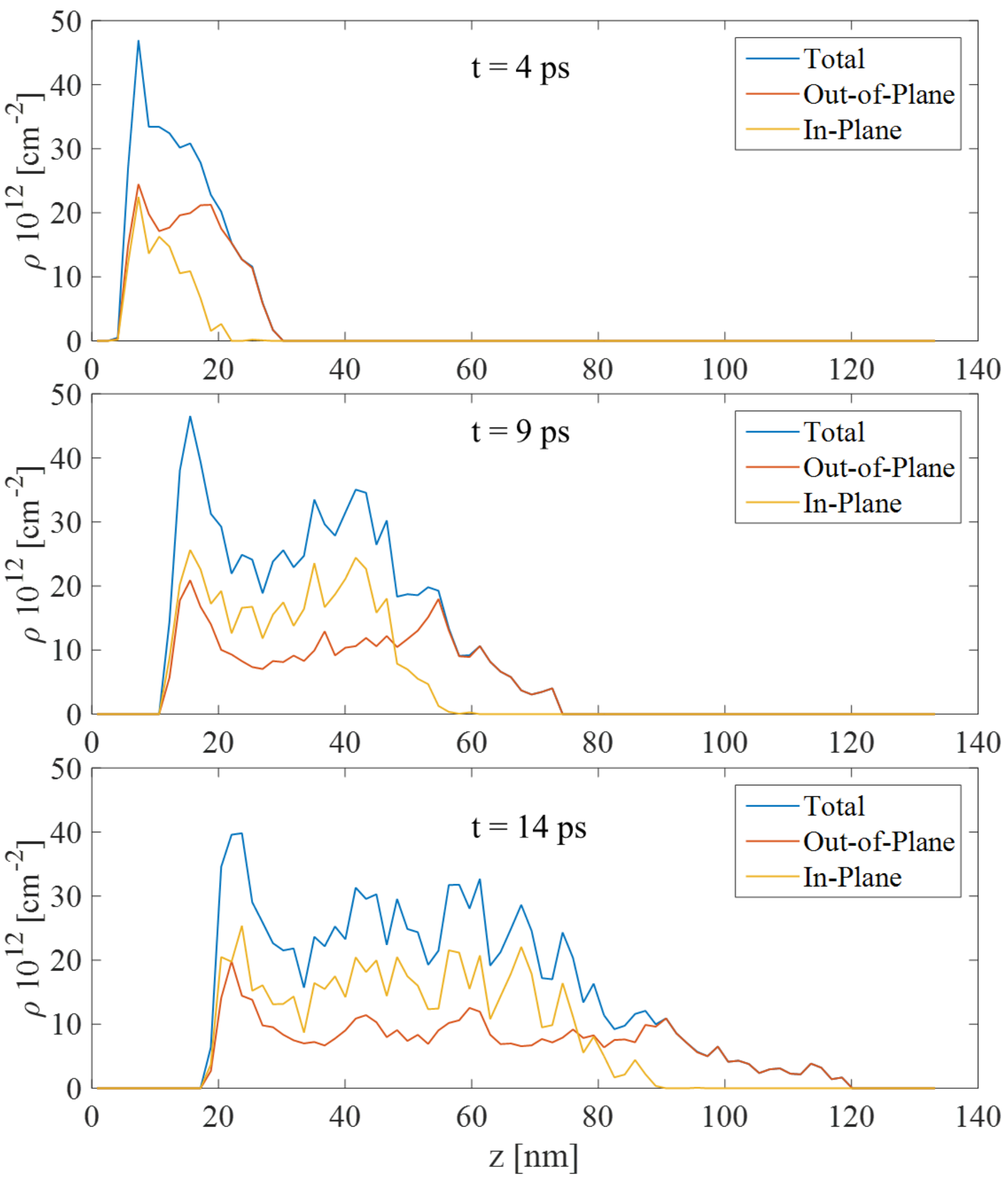

**Figure 6 - Dislocation density profiles at 4, 9 and 14 ps from loading in a single simulation**

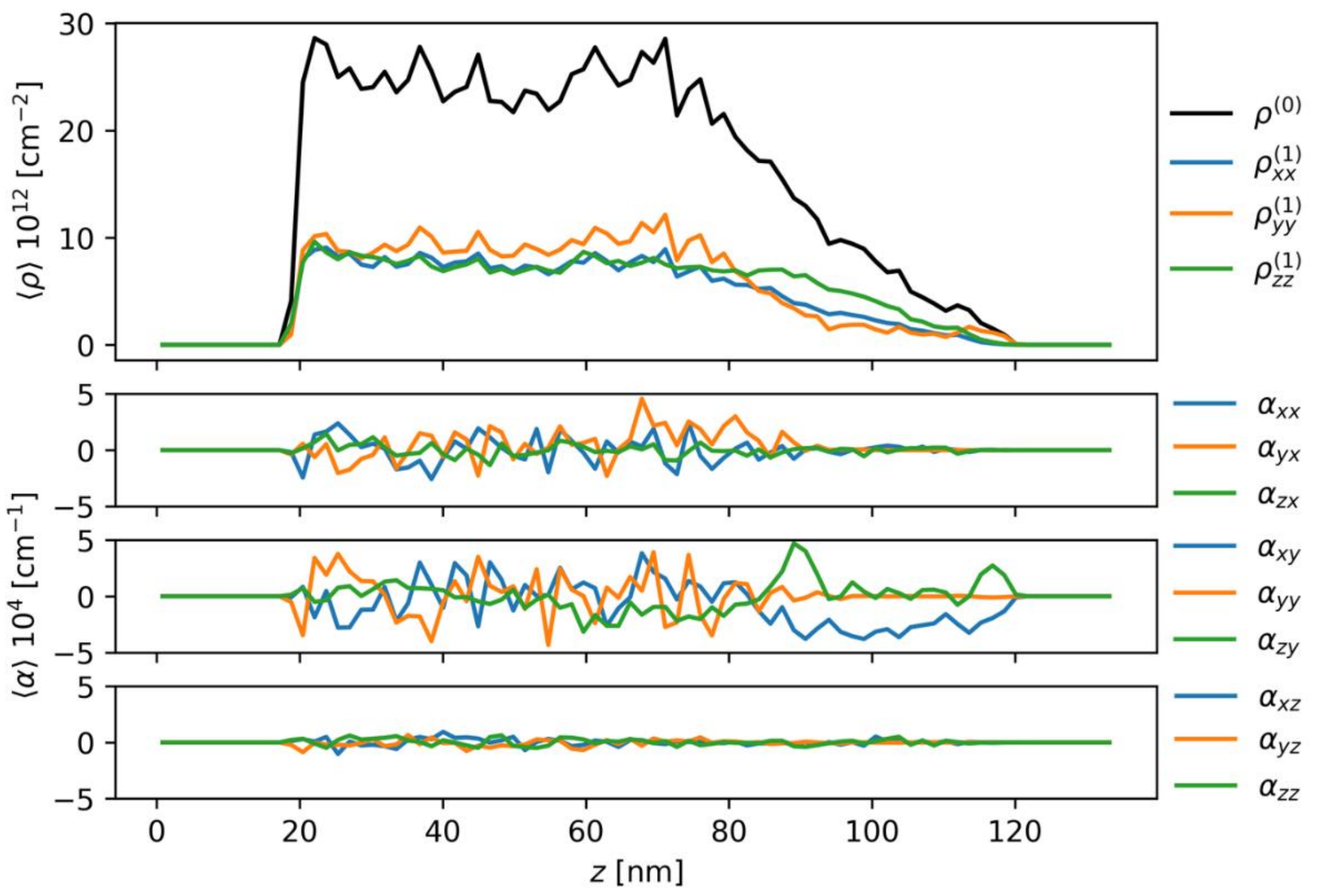

**Figure 7 - Ensemble averages of different density types and Kröner–Nye tensor components at time $t = 14$ ps.**

# 4. Discussion

## 4.1. Dislocation evolution and resolved shear stresses

The dislocation densities can be directly correlated to the stress profiles, as we demonstrate in Figure 8 at the time of 14 ps after initial loading. At this time, the front of the stress wave has traveled 128 nm from the impact surface while the front of the leading dislocation is located at 120 nm. Thus, we can say that a short elastic precursor wave is observed. Following the elastic precursor, we observed a gradual increase in the out-of-plane dislocations without formation of any in-plane dislocations. The out-of-plane dislocations nucleate and dislocation lines expand until reaching a density of about $1\times10^{13}$ cm$^{-2}$. The density composition with respect to the contributing line directions is in good agreement with the argument that more mobile edge dislocations leave behind stems of long screw segments. This is supported by the Kröner-Nye tensor components in the shock front, as the ones corresponding to edge character ($xy$ and $zy$) are the only ones that are non-zero in this regime. A stronger influence of the screw components with $z$-direction line component is partially cancelled out by the left behind stems which always form pairs of opposite line direction. As the dislocations within the shock front are of out-of-plane type the $x$-line-direction-contribution is negligible. At this point, we observe the onset of nucleation of the in-plane dislocations. As these dislocations nucleate, a rapid increase in $\sigma_y$ is seen and, as a result, the in-plane shear stresses are reduced. The in-plane dislocation density reaches a saturated value of about $1.5\times10^{13}$ cm$^{-2}$ at about 55 nm behind the shock front. Finally, we obtain a total dislocation density of about $2.5\times10^{13}$ cm$^{-2}$ which is in good agreement with Meyers homogenous nucleation model [15]. This density is of mostly SSD type, which may be explained by the governing hydrostatic stress which does not require excess densities in order to counter the elastic stresses.

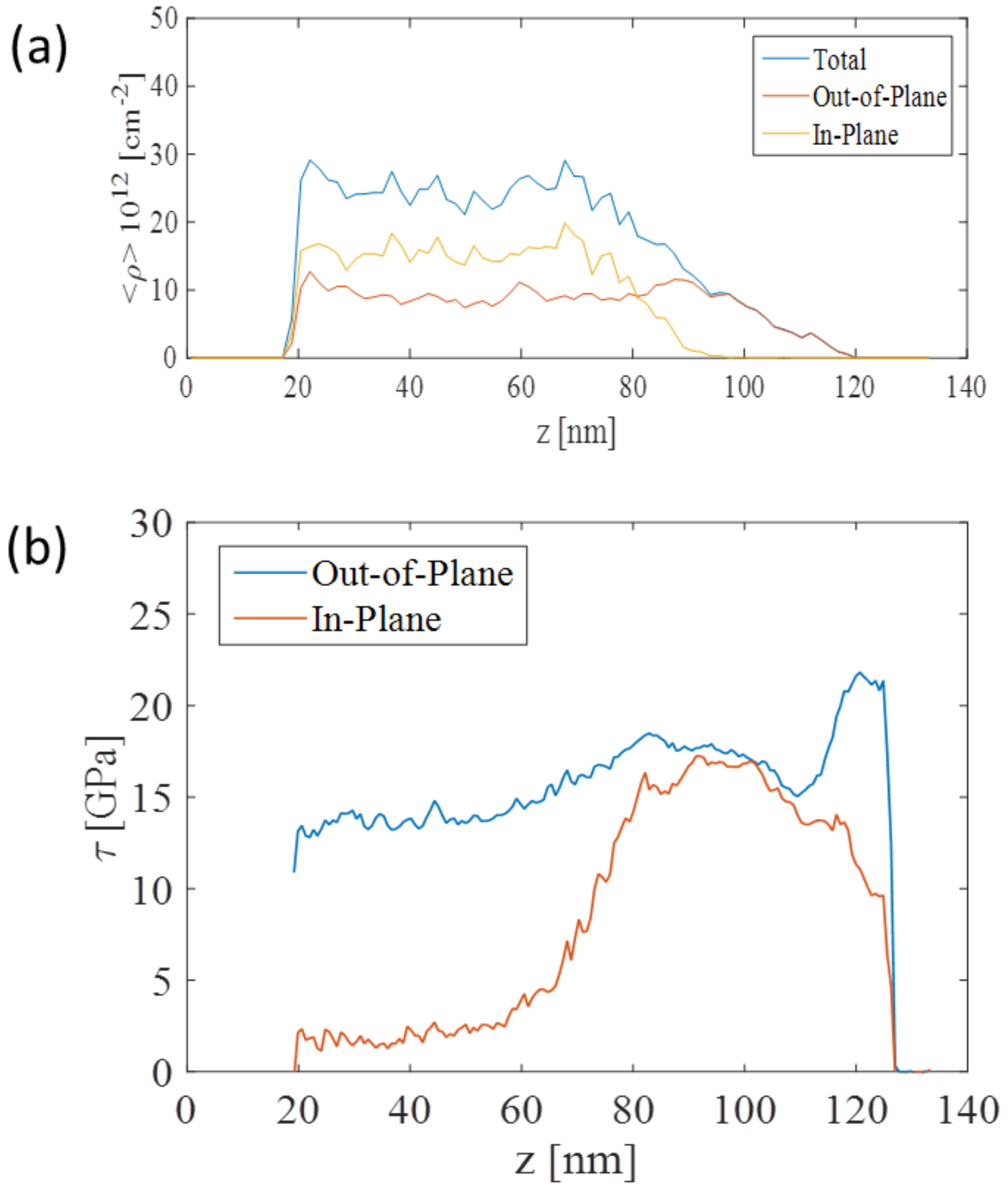

**Figure 8 – The correlation between dislocation densities and the local stress. (a) Average dislocation profiles from 5 simulations at 14 ps after loading. (b) Resolved shear stress in the out-of-plane and in-plane directions at the same time.**

Knowing the total stress tensor, we can calculate the resolved shear stress (RSS) in the out-of-plane and in-plane directions. The RSS profiles at 14 ps in out-of-plane and in-plane directions are drawn in Figure 8b. We initially see a sharp jump of the RSS in the out-of-plane direction to a value of about 21 GPa. This value is higher than 15 GPa, which is the theoretical shear stress of Mo found in [45]. Despite, this stress level is maintained for a short duration (seen as a constant stress in the profile) and then starts to decrease in correlation with the nucleation of the out-of-plane dislocations. The nucleation of dislocations relives the RSS in the out-of-plane direction until it reaches a constant level of about 14.5 GPa. This result is a reminiscence of the stress overshoot in amorphous solids, in which the stress is increased above the limit to trigger shear transformation [46]. Shear transformation, similarity to nucleation, is a thermally activated process, and we propose that the elastic region duration and the stress overshoot depend on the response of the pristine crystal to nucleation dislocations.

As for the RSS in the in-plane direction, it jumps initially to a value of about 10 GPa which is below the nucleation stress found in [45] and thus insufficient for dislocation nucleation. For reasons that will be further clarified, the RSS in the in-plane direction rises until it reaches about 16 GPa which is close to nucleation stress. As the rate of increase in the in-plane dislocations decreases, the shear stress drops and reaches steady state at a value of about 2 GPa. It is commonly accepted that dislocation nucleation and glide results in plastic strain and the reduction of shear stresses. This concept is the basis of many continuum mechanics plasticity models. However, we identified in our simulations that the reduction in the shear stresses is accompanied by an increase in the principal stresses and, thus, the deviatoric and volumetric response is coupled. While others [47] have also reported on the evolution of principal stresses, we wish to dwell on this.

The stress evolution inside a lattice is controlled by the elastic strains that develop during deformation. However, it is not the elastic strains that are prescribed by the boundary conditions but the total strains, since this system is displacement-driven. For the case of shock compression, we prescribe the lateral *total* strains to be zero. Therefore, if plastic strain evolves due to some micromechanical mechanisms, it must be accompanied with an opposite elastic strain which will result in an increase in stress. An example of this type of analysis can be seen in [5] where the strain in the shocked sample was measured *in-situ* using x-ray diffraction. In our case, the high stress behind the shock front induces dislocation nucleation and glide in the out-of-plane directions, and later in the in-plane directions, which results in a resolved plastic strain on the different plane slip planes. Thus, there are non-zero plastic strain components in the principal directions, which results with non-zero elastic strain component that nullifies the total lateral strain, i.e., lateral compressive stresses develop in the lattice. We better quantify the strains in the following section.

## 4.2. Analysis of strains behind the shock front

In order to quantify the plastic strains that evolve behind the shock front we follow a similar analysis to the one performed in [5]. The total strain at a material point is

$$\boldsymbol{\varepsilon}_{total}(z) = \boldsymbol{\varepsilon}_E(z) + \boldsymbol{\varepsilon}_T(z) + \boldsymbol{\varepsilon}_p(z). \tag{4}$$

where $\boldsymbol{\varepsilon}_E, \boldsymbol{\varepsilon}_T, \boldsymbol{\varepsilon}_p$ are the average elastic, thermal and plastic strain tensors at a certain slab with a common $z$ coordinate, respectively. From the shock simulations presented above, we obtain the atomic temperature and compressive axial strain in the $z$ direction behind the shock. Based on the change in density, the total uniaxial compressive strain in the shock direction reaches a value of about 0.22 in the steady-state regime, far from the shock front. We recall that in this work we refer to compressive strains as positive.

The contribution of the thermal strain can also be estimated from the simulations. The temperature of the compressed material behind the shock rises by about 1500 K. To estimate the accompanying thermal stress at this temperature, we performed specific MD simulations to calculate the thermal expansion in equilibrium. The MD simulation computational cell is fully periodic (which represents an infinite bulk), coupled to the Nosé-Hoover thermostat and the Parrinello-Rahman barostat, to allow heating the system while relaxing the shape. For a temperature rise $\Delta\mathrm{T} = 1500$ K we got about -0.01 strain (negative for expansion), which results in thermal expansion coefficient similar to the experimentally measured thermal expansion coefficient for Mo [48]. In addition to the thermal strains, we should also calculate the elastic strains contributions to the total strain. To relate directly the elastic strains to the interatomic potential properties at high stresses, we performed a MD simulation of a bulk crystal, in the same orientation as the impacted thin-film, using constant temperature and stress conditions, with a prescribed external stress with $\sigma_x$, $\sigma_y$ and $\sigma_z$ equal to 104 GPa, 108 GPa and 140 GPa, respectively, as obtained in the steady-state regime behind the shock front. Despite the high stresses, we emphasize that the lattice remained elastic during the simulation. The compressive elastic strains extracted from the MD simulations, which correspond to the stresses stated above, are 0.06, 0.04 and 0.13 in the $x, y$ and $z$ direction, respectively. Knowing the total strains and using Eq. (8) we can calculate the plastic strains in the three principle directions: -0.05, -0.03 and 0.10 in the $x, y$ and $z$ direction, respectively.

The plastic strains are based on the dislocation microstructure evolution, through Orowan's equation:

$$\gamma = \rho b \Delta x, \tag{5}$$

where $\gamma$ is the resolved plastic strain in the slip direction, $\rho$ is the corresponding dislocation density, $b$ is the Burgers vector and $\Delta x$ is the mean free path of dislocations. We note that since dislocation are both nucleating and gliding behind the shock front, we did not use the commonly used relation for the plastic strain rate. The expression for the plastic strain rate would have involve both the mean velocity, instead of the mean free path, and the nucleation rate, both of which we do not know. The resolved plastic strain on a specific plane in a specific direction can be calculated form the plastic strain tensor,

$$\gamma = (\boldsymbol{\varepsilon_p} \cdot \boldsymbol{n}) \cdot \boldsymbol{s}, \tag{6}$$

where $\boldsymbol{n}$ is the normal to the slip plane and $\boldsymbol{s}$ is the slip direction. The value of $b$ is known and $\rho$ can be extracted from the dislocation density analysis we have performed. Taking the results of plastic strains obtained in the first part of this section and combining Eq. 9 and 10, we can approximate the values of $\Delta x$ on each slip system. For the in-plane dislocations taking $b = 0.285$ nm and $\rho = 0.75 \cdot 10^{13}$ cm$^{-2}$ we get $\Delta x = 0.4$ nm. For the out-of-plane dislocations, taking the same $b$ as previously and $\rho = 0.5 \cdot 10^{13}$ cm$^{-2}$ we get a free mean path of $\Delta x = 4.8$ nm. We note that as there are two slip directions in the in-plane and out-of-plane directions, we have to take the total dislocation density for the respective direction. The calculated results serve as a rough estimate of the mean free path. The calculated values of $\Delta x$ for both the in-plane and out-of-plane directions are low, with respect to values on the mean free path reported in literature. For instance, Devincre et al. calculated the mean free path of dislocations in Cu using discrete dislocation dynamics (DDD) simulations [49], and found it to be of the order of $10^3 - 10^4$ nm for an initial dislocation density of $2 \cdot 10^8$ cm$^{-2}$. Given that the dislocation densities calculated behind the shock are 5 orders of magnitude are higher than the values prescribed in the DDD simulations, the mean free path of dislocations behind the shock is expected to decrease as the inverse of the square root of the dislocation density. Thus, values of few nm to few tens of nm are expected. These low values are a testament to the significant contribution of homogenous nucleation, rather than glide, to the plastic strain in shocked single crystals. The very high dislocation densities obtained in MD and the short glide distance should be taken into account when modeling plasticity at these extreme pressures and strain rates. Thus, flow rules defined solely based on dislocation glide can be unappropriated and dislocation nucleation should be taken into account [50].

## 5. Summary

The MD simulations, in combination with the D2C method, allowed us to quantify the stress relaxation during shock-induced plasticity in Mo impacted along the <110> direction. The elastic precursor was found to overshoot the compressive stress, which is relaxed slightly behind the elastic front by nucleating dislocations in the out-of-plane directions. The dislocations contribution is twofold, on the one hand to relax the shear stresses in the out-of-plane direction, but at the same time increasing one of the lateral principal stresses. As a result, resolved shear stresses in the plane parallel to the shock front increases and a second front of dislocation nucleation in the in-plane direction lags behind. The plastic front eventually leads to an isotropic stress condition in the plane parallel to shock front.

We note that this result is not specific for Mo, but depends on the crystal orientation. Thus, the conclusions above are relevant for all BCC metals, as long as the plastic deformation is dislocation-mediated. The contribution of solid-solid shock-induced phase transformation is left for future work. We also note that the two-stage plastic deformation is particular to this crystal orientation and impact in different directions (e.g., <100> or <111>) may nucleate dislocations on different slip planes simultaneously. Despite, the analysis proposed here demonstrated the importance of having a quantitative technique to analyze dislocation evolution on particular slip planes from MD simulations and how to relate them to continuous fields.

## 6. Acknowledgements

DS and SS acknowledge support from a European Research Council Starting Grant, “A Multiscale Dislocation Language for Data-Driven Materials Science”, ERC Grant Agreement No. 759419 MuDiLingo. RK and DM wish the thank Prof. Eugene Zaretsky for stimulating discussions.